\documentclass[preprint,12pt]{elsarticle}

\usepackage{amssymb}

\usepackage{gensymb}
\usepackage[english]{babel}
\usepackage{microtype}
\usepackage{booktabs}
\usepackage{doi}

\usepackage{amsmath}

\journal{Applied Surface Science}

\begin{document}

\begin{frontmatter}



\title{Pulsed laser synthesis of mesoporous metal chalcogenide thin films}


\author[inst1]{Dorien E. Carpenter}
\author[inst1]{Zahra Nasiri}
\author[inst1]{Nithesh R. Palagiri}
\author[inst1]{Kamron L. Strickland}
\author[inst2]{Sumner B. Harris}
\author[inst3]{David B. Geohegan}
\author[inst1]{Renato P. Camata\corref{cor1}}
\ead{camata@uab.edu}
\cortext[cor1]{Corresponsing author}

\affiliation[inst1]{organization={Department of Physics,\\ University of Alabama at Birmingham},
            city={Birmingham},
            state={Alabama},
            postcode={35294}, 
            country={USA}}

\affiliation[inst2]{organization={Center for Nanophase Materials Sciences,\\ Oak Ridge National Laboratory},
            city={Oak Ridge},
            state={Tennessee},
            postcode={37831}, 
            country={USA}}

\affiliation[inst3]{organization={Department of Materials Science and Engineering,\\ University of Tennessee at Knoxville},
            city={Knoxville},
            state={Tennessee},
            postcode={37996}, 
            country={USA}}

\begin{abstract}
Mesoporous films of the metal chalcogenide $\beta$-FeSe were grown on MgO substrates by KrF pulsed laser deposition (PLD) in an argon background. At 100 mTorr, gated intensified charge-coupled device imaging and ion probe measurements showed that the plasma plume responsible for crystal growth initially comprised three components, with distinct expansion velocities. Plume interactions with the substrate heater and ablation target gave rise to complex dynamics, including collisions between the charged leading edge---rebounding between the substrate and the target---and slower-moving species in the plume interior. Film growth was dominated by species with kinetic energies $\le$0.5~eV/atom. X-ray reflectivity and atomic force microscopy revealed that films grown in this environment---with a substrate temperature of 350$\degree$C, a laser fluence of 1.0 J cm$^{-2}$, and a 7.5 mm$^2$ spot area---formed a porous framework with 15\% porosity and pore sizes below 100 nm. X-ray diffraction indicated that the porous films were epitaxial with respect to the substrate and likely grew by oriented-attachment of gas-phase molecular clusters or very small nanoparticles, in contrast to the conventional epitaxy of vacuum films from atomic constituents. The in-plane orientation of the mesoporous films was $\beta$-FeSe~[100]$\parallel$[110]~MgO, attributed to the soft landing of pre-formed crystallites on the MgO substrates, where protruding Se rows of $\beta$-FeSe aligned with corrugations of the MgO surface. This work implies that growth of candidate electrocatalyst materials by PLD in inert gas background may allow mesoporous frameworks with a single crystallographic orientation that expose specific crystal facets for electrochemical reactions and active site engineering.

\medskip\noindent
    \footnotesize{Notice: This manuscript has been authored by UT‑Battelle, LLC, under Contract No.\ DE‑AC05‑00OR22725 with the U.S.\ Department of Energy. The United States Government retains and the publisher, by accepting the article for publication, acknowledges that the United States Government retains a non‑exclusive, paid‑up, irrevocable, world‑wide license to publish or reproduce the published form of this manuscript, or allow others to do so, for United States Government purposes. The Department of Energy will provide public access to these results of federally sponsored research in accordance with the DOE Public Access Plan (http://energy.gov/downloads/doe-public-access-plan).}

\end{abstract}

\begin{graphicalabstract}
\includegraphics[width=\linewidth]{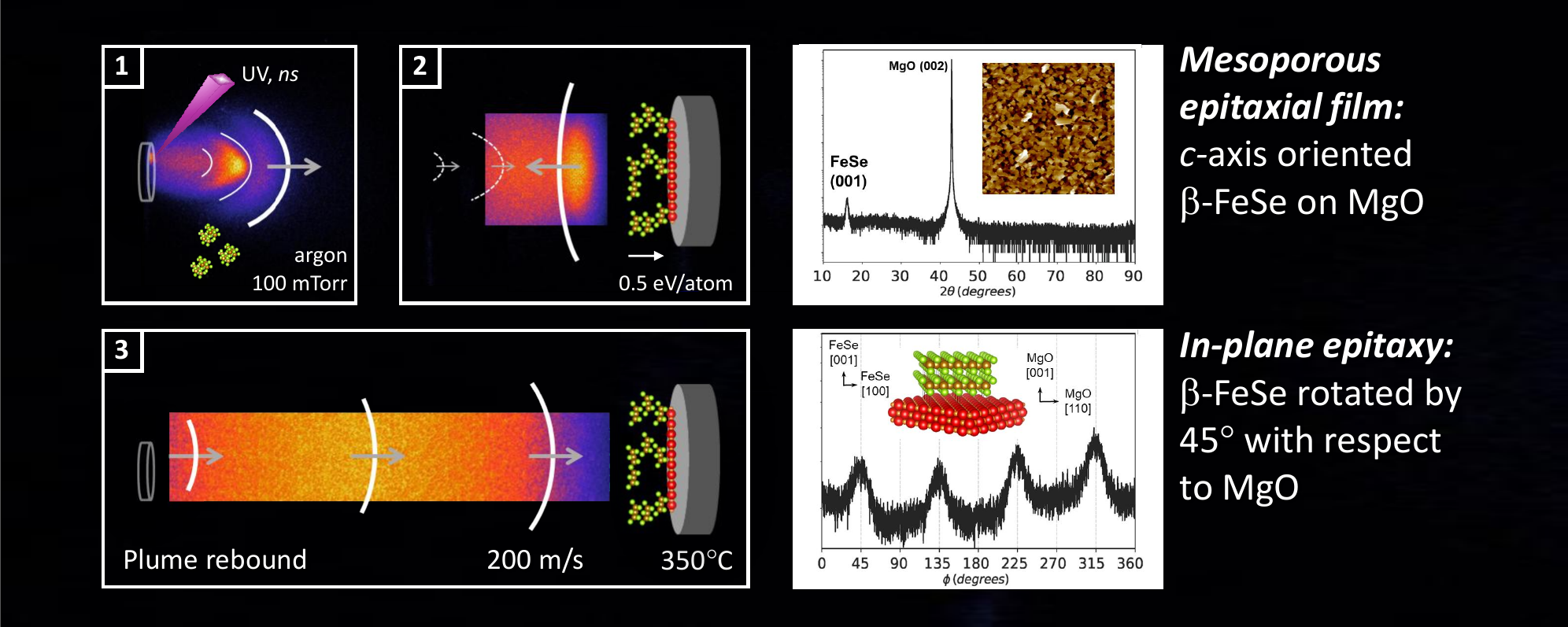}
\end{graphicalabstract}

\begin{highlights}
\item Mesoporous metal chalcogenide $\beta$-FeSe films grown on MgO by pulsed laser deposition
\item Films are epitaxial with $c$-axis of $\beta$-FeSe oriented normal to the MgO substrate
\item In-plane, the $\beta$-FeSe unit cell is rotated by 45$\degree$ relative to the MgO lattice
\item The cluster- or nanoparticle-based films arise amid complex plasma plume dynamics
\item Plume interactions expand options in metal chalcogenide electrocatalyst engineering
\end{highlights}

\begin{keyword}
metal chalcogenide \sep cluster epitaxy \sep oriented attachment \sep laser-generated plasma \sep mesoporous \sep plume dynamics \sep electrocatalyst


\end{keyword}

\end{frontmatter}



\section{Introduction}
\label{sec1}
The development of high-performance electrocatalysts is important for increasing the efficiency of hydrogen production by water electrolysis. Designing stable electrodes from earth-abundant materials, capable of promoting high rates of the electrocatalytic hydrogen evolution reaction (HER) and oxygen evolution reaction (OER), could bring about broader adoption of water-splitting technologies. Among the many materials being explored for this purpose, metal chalcogenides are some of the most promising~\cite{zhao2021metal,Majhi2023}. The electronic properties of these compounds, including their band gap and electron mobility, can be controlled through phase and compositional tuning~\citep{Woods-Robinson2020,Zhou2023,Giri2023}. Their complex doping effects and surface reconstruction patterns provide rich opportunities for active site engineering~\citep{Fabbri2017,Wang2020}. In certain configurations, they may also enable photo-enhanced catalytic processes~\cite{ha2023recent}. Together, these features represent a compelling platform for the development of efficient electrocatalysts. Some metal chalcogenides have shown performance that matches or exceeds the record function of the platinum group metals (Pt, Pd, Rh) for HER and oxides (IrO$_2$, RuO$_2$) for OER~\cite{ha2023recent}. MoS$_2$, for example, excels in HER~\cite{sun2017enabling}, while Ni$_3$Te$_2$ outperforms RuO$_2$ and IrO$_2$ in OER~\cite{de2018nickel}. FeSe$_2$ and CoSe$_2$ are also noteworthy as potential bifunctional catalysts for overall water electrolysis in alkaline conditions~\cite{panda2017molecular,lu2018reaction}. Moreover, recent studies show that efficiency gains can be achieved when metal chalcogenides are used in hybrid configurations with other materials and heterostructures~\cite{zhao2019nico,peng2020emerging,luo2020interface,Li2023}. These encouraging results are driving studies to fine-tune metal chalcogenides at atomic and mesoscale levels. Synthesis routes that allow porous architectures and exposure of specific crystal facets are particularly desirable. They may enable electrodes with large surface areas, high pore volumes, and uniform mesopore shapes, while exposing active crystal facets that are optimal for adsorption and interaction with reactants~\cite{Miao2017,Tufa2021}. In this work, we explore the pulsed laser synthesis of thin films of the metal chalcogenide iron selenide (FeSe), with potential for electrocatalytic applications due to their mesoporosity and preferred crystallographic texture.

When the ejecta produced by laser ablation of a solid collides with an inert gas background, the expanding plume decelerates, cools, and becomes confined. The resulting material flux is typically dominated by molecular clusters and larger particles. Films grown from these objects can develop a porous framework if the substrate temperature is insufficient to induce their melting and re-crystallization~\citep{Jensen1999}. This approach has been used to grow mesoporous thin films of a variety of materials, including yttrium-stabilized zirconia~\cite{Nair2005}, tungsten oxide~\cite{di2006synthesis} and trioxide~\cite{shin2015tree}, titanium dioxide~\cite{sauvage2010hierarchical,  noh2012aligned}, and carbon foams~\cite{maffini2022pulsed}. The method allows control of crystallographic orientation, grain size, and porosity. In addition, it is compatible with exploration of diverse chemical species. Its versatility also makes it suitable for investigating doping effects, phase optimization, heterostructure formation, and the development of hybrid materials. We apply this methodology to the synthesis of porous FeSe thin films. A gated intensified charge-coupled device (ICCD) camera and an ion probe are used to monitor the expansion of the laser-generated plasma plume. The microstructure, porosity, and crystallographic structure of the films are evaluated using X-ray reflectivity (XRR), X-ray diffraction (XRD), and atomic force microscopy (AFM).

\section{Experimental details}
\label{sec2}
A KrF excimer laser (Coherent LPX Pro; 248 nm wavelength; 25 ns pulse duration) was used to ablate a rotating metal chalcogenide target inside a turbomolecular-pumped vacuum chamber ($1.2 \times 10^{-6}$ Torr base pressure). The target, in the shape of a \mbox{1-inch} diameter disk and with a nominal stoichiometry of Fe$_{1.03}$Se, was synthesized by solid state reaction from elemental metal powders, pressed, and sintered according to the standard procedures described in Ref.~\citenum{feng2018tunable}. During ablation, a 20--100 mTorr inert background was produced by flowing argon (Ar) gas into the chamber at a rate of 65 sccm while adjusting the pumping speed of the vacuum system. The irradiation spot on the target was obtained by imaging a rectangular aperture placed in the beam path between the laser output window and the focusing lens. The aperture was sized to generate a spot area of 7.5~mm$^{2}$ (3.0~mm~$\times$~2.5~mm) on the target surface and the energy of the laser pulse was adjusted to deliver a fluence of 1.0 J cm$^{-2}$ over the irradiated area. Further details of the experimental configuration have been described elsewhere~\cite{harris2019double, Harris2019b}.

The expansion behavior of the ablation plume was observed using ICCD photography of the visible laser-generated plasma intensity. Sequences of 20--40 ICCD images (16-bit grayscale) were collected for single-shot laser pulses with delay times from 0.5 $\mu$s to 5 ms and the gate time set to 10\% of the delay time for each image. The ICCD camera (Princeton Instruments, PI-MAX 4) was positioned 80 cm away from the center of the plume and used an f2.8 camera lens (105 mm micro-NIKKOR). The \texttt{CMRmap} colormap, available in the Python-based Matplotlib visualization library~\cite{Hunter2007}, was used to display the intensity data. Ion probe waveforms were measured using a biased wire placed in front of the substrate ($-40$ V bias, 4 mm long, 0.4 mm diameter) and an oscilloscope (1 GHz, Tektronix MSO64) with a 50 $\Omega$ feed-through BNC resistor. The ion probe was positioned at 4.8 cm from the target along the symmetry axis of the plume. It was mounted on a rotary vacuum feedthrough, which also supported a metallic shield placed 0.36 cm behind the probe, to protect the substrate during probe measurements. For film deposition, the probe and shield were retracted, and the substrate heater was moved to approximately the same position formerly occupied by the shield. During ICCD imaging and ion probe data acquisition, the substrate shield and the receded heater were kept at room temperature. Both the ICCD camera and oscilloscope were triggered by a photodiode monitoring the KrF laser pulses.

Films were grown on high-quality (100)-oriented MgO substrates free from  crystal twinning (MSE Supplies LLC, Tucson, Arizona, USA). To confirm that all substrates were single crystal, a rocking curve scan was acquired around the (0 0 2) MgO peak. Only substrates showing a single well-defined peak, without evidence of peak splitting due to twinning, were selected. Substrates were placed perpendicularly to the symmetry axis of the ablation plume, at a distance of 5.0 cm from the target. Prior to deposition, substrates were cleaned by agitation in a Hellmanex III detergent solution (Helma GmbH, M\"ullheim, Germany), followed by a series of deionized water (5 min for 7 times), acetone (10 min) and methanol (10 min) rinses in an ultrasonic bath. Upon removal from the final rinse, each substrate was blown dry with nitrogen and promptly placed in the vacuum chamber. This method of substrate preparation removes any Mg(OH)$_2$ formed due to ambient exposure and also eliminates other surface contaminants including hydrocarbons. Before each deposition, the target was pre-ablated with 1,000 pulses while the clean substrate was shielded from the plume to ensure removal of any oxides or other surface impurities. Films were grown at a substrate temperature of 350$\degree$C using 6,000 KrF pulses with a repetition rate of 5 Hz.

The surface morphology of the resulting films was analyzed using a Bruker Dimension Icon AFM in tapping mode with a Si probe (TESPA-V2, 7 nm tip radius, 37 N/m spring constant). X-ray reflectivity (XRR) measurements were performed using a high-resolution X’Pert Pro MRD diffractometer (Malvern Panalytical) equipped with a Hybrid 4-bounce Ge (220) monochromator providing Cu K$\alpha_1$ radiation ($\lambda = 1.5406$ Å). A 1/32$\degree$ divergence slit was used to define the incident beam. The reflected intensity was recorded as a function of the incident angle over the range from 0.03$\degree$ to 1.5$\degree$, with a step size of 0.005$\degree$. 

To determine thin film parameters such as mass density, porosity, and thickness, a three-layer model comprising the MgO substrate, the FeSe film, and air (above the sample) was fitted to the XRR data. The model, which computes the X-ray reflectivity for the three-layer system recursively using the Fresnel reflection coefficients at each interface~\citep{Parratt1954}, is implemented in the open-source GenX 3 software~\cite{Glavic2022}. GenX 3 utilizes differential evolution refinement to determine the best model parameters. 

X-ray diffraction (XRD) measurements were conducted on the X'Pert Pro MRD system mentioned earlier ($\theta$-2$\theta$ scans) and also on a high-resolution Empyrean, Malvern Panalytical multipurpose diffractometer ($\phi$ scans). Three-axis goniometers were used in both diffractometers.

\section{Results and discussion}
\label{sec3}

\begin{figure*}[!t]
    \centering
    \includegraphics[width=0.95\linewidth]{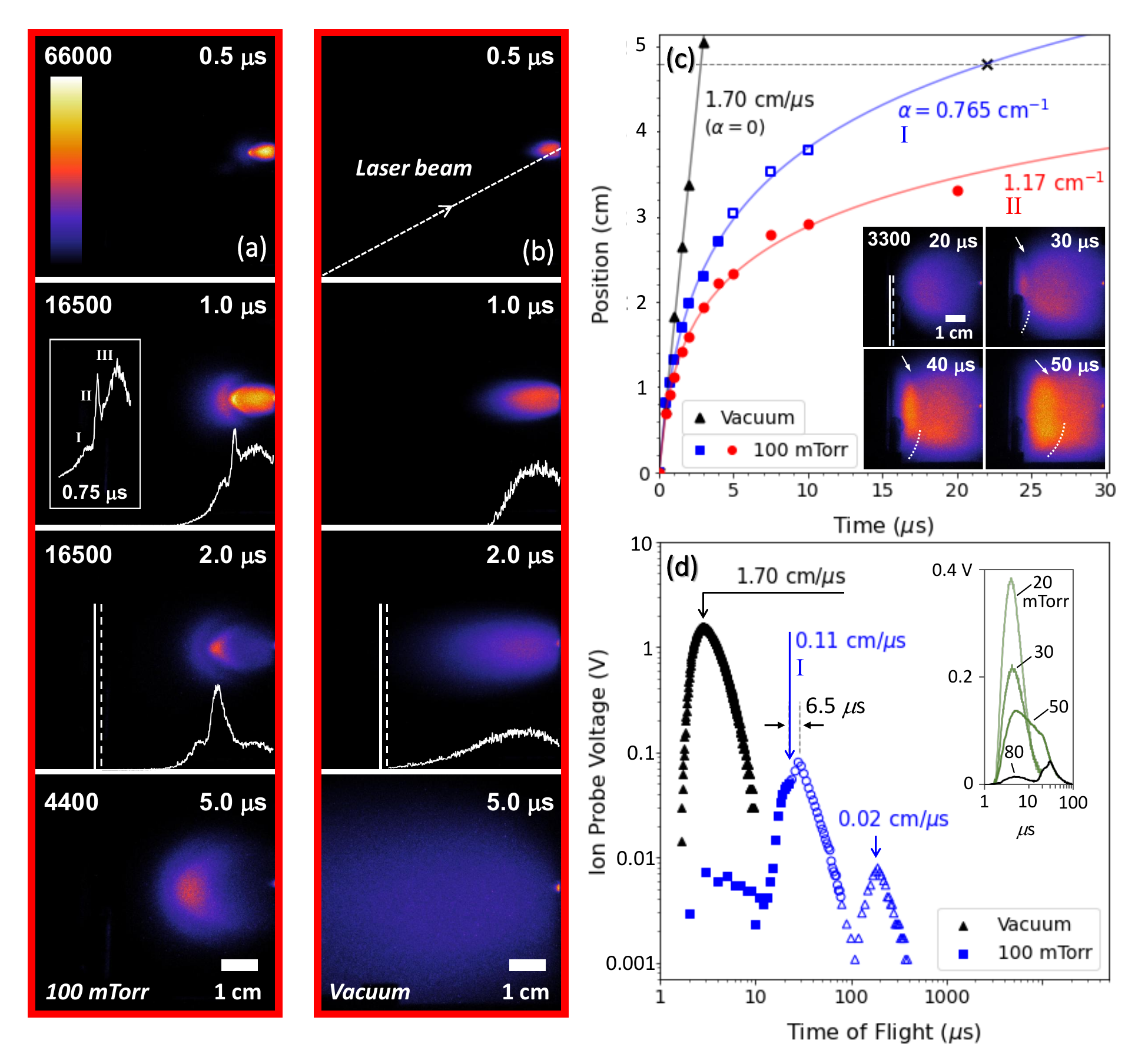}
    \caption{Gated ICCD images of the visible plume following single-shot ablation of FeSe, shown for (a) 100 mTorr Ar and (b) vacuum ($1.2 \times 10^{-6}$ Torr) at representative delay times. Insets in the 1.0 and 2.0~$\mu$s panels show intensity vs. position profiles along the plume’s symmetry axis; the 0.75~$\mu$s profile highlights early components (I, II, III). The maximum intensity used to rescale the false-color palette appears in the top-left of each 100 mTorr image; corresponding vacuum images are scaled identically. Vertical dashed (solid) lines mark the ion probe (shield/substrate) positions. (c) $R$–$t$ plots of select plume structures. “X”-mark on the horizontal dashed line (probe location) indicates arrival time of feature I, which matches a quadratic drag model of plume expansion ($a = -\alpha v^2$, with $R = \alpha^{-1} \ln(1 + \alpha v_0 t)$). Solid symbols mark features clearly identified in ICCD profiles; open symbols correspond to lower-confidence assignments. As $t \to 0$, $v_0$ of background expansion approaches the vacuum value of 1.70 cm $\mu$s$^{-1}$. Inset shows plume front reflected from the shield and its interaction with II. (d) Ion probe signal for vacuum and 100~mTorr expansions. Blue labels indicate probe-arrival velocities of front species for the corresponding structures. Open circles and triangles denote reflected and rebounded front, respectively. Inset shows how increasing pressure alters the waveform. ICCD and probe data collected with the probe and shield inserted, and the substrate heater receded and maintained at room temperature.
    }
    \label{fig:ICCD}
\end{figure*}

Figure \ref{fig:ICCD}(a) shows gated ICCD images of the visible plume for representative delay times of 0.5, 1.0, 2.0, and 5.0 $\mu$s after single-shot ablation of the FeSe target into 100 mTorr Ar background (see Fig. S1 and Movie S6 in Supporting Information for images at longer delay times). For comparison, Fig. \ref{fig:ICCD}(b) displays images for an expansion into vacuum under the same laser irradiation conditions. Consistent with numerous previous studies, the expansion into inert background shows a bow shock structure and splitting of the visible plume into several components. The leading edge, labeled ``I'' in the inset of the 1.0 $\mu$s panel, is characterized by a sharp intensity front marking the outermost extent of the luminous emission. A separate high-intensity feature, resembling an arrowhead in shape, trails the leading edge. This feature, labeled ``II'' in the inset, is particularly clear in the 2.0~$\mu s$ image. A third, broader visible emission structure (III) lags behind and rapidly decays in intensity. The presence of such complex flow pattern stands in contrast to the simpler deflagration behavior of the expansion into vacuum. The significant deceleration of the visible plume in the Ar background is apparent for $t > 1 \: \mu \text{s}$. At 5 $\mu$s, for example, the front expanding into background is still at 3 cm from the target, while the expansion in vacuum has already filled the entire field of view of the camera, which measures 6.6~cm across. Plots of the position of plume structures versus time ($R$-$t$), extracted from ICCD images, are shown in Fig. \ref{fig:ICCD}(c). In vacuum, the velocity of the expansion front is constant, while the expansion into the inert background is characterized by decreasing velocities.

Ion probe measurements detect charged plume species beyond the range of the luminous emission captured by photography. Combined analysis of ICCD exposures and ion probe data permits a better-informed interpretation of plume dynamics by correlating optical emission features with ion transport behavior. The ion probe data for our expansions are shown in Fig. \ref{fig:ICCD}(d). A single peak is observed for the expansion into vacuum. Ions giving rise to this peak travel at a centroid speed of 1.70 cm~$\mu$s$^{-1}$ (17.0~km~s$^{-1}$). Invoking an ion mass of $1.12\times 10^{-25}$~kg (average of Fe and Se), this high speed corresponds to a kinetic energy of 101 eV/atom. On the other hand, four distinct features are observed in the ion probe signal for the background expansion. The 1.70~cm~$\mu$s$^{-1}$ peak is strongly attenuated to less than 1\% of its vacuum intensity. Additional peaks, associated with significantly longer times of flight, now dominate the probe signal, indicating the presence of slower-moving objects in the plume. A double-peak structure is detected in the \mbox{20--30~$\mu$s} time-of-flight range, with overlapping ion peaks separated by 6.5~$\mu$s. 
Arrival at the probe of a charged population with time of flight of 186~$\mu$s is also observed. The inset in Fig. \ref{fig:ICCD}(d) shows how the single-peaked vacuum waveform evolves into the multiple peak structure in the Ar background, as the pressure is raised through 20, 30, 50, and 80 mTorr.

Generally speaking, the splitting of the ablation plume into several velocity components can be understood in terms of scattering events between atoms ejected from the target and background species, with scattering cross sections of the order of atomic values~\cite{Wood1998a, Wood1998b}. In this scattering scenario, distinct peaks of the ion probe signal may originate from composite distributions of target atoms that experience discrete numbers of collisions as they traverse the distance between target and probe~\cite{Wood1998a, Wood1998b}. The residual fast peak (1.70~cm~$\mu$s$^{-1}$) in the 100 mTorr probe signal, for example, is attributed to material that penetrates the background gas without undergoing significant scattering. 

A joint examination of Figs. \ref{fig:ICCD}(c) and (d) reveals additional plume interaction effects that contribute to the ion probe waveform observed at 100 mTorr. It is worth noting initially that the probe-derived centroid velocity of ions in vacuum, shown in Fig. \ref{fig:ICCD}(d), matches the speed of the leading edge of the ICCD images, plotted in Fig. \ref{fig:ICCD}(c), provided the vacuum leading edge is defined as the position beyond which the plume emits 7\% of the total integrated intensity. The leading shoulder of the 100 mTorr double peak \mbox{(20--30~$\mu$s)} arrives at the probe at 22~$\mu$s. It is represented by an ``x-mark'' in Fig. \ref{fig:ICCD}(c). This probe signal is consistent with the time evolution of the ICCD leading edge (I) according to a quadratic drag model, $R = \alpha^{-1} \mathrm{ln}(1+\alpha v_0t)$, with $\mathrm{\alpha = 0.765\ cm^{-1}}$, as shown in Fig. \ref{fig:ICCD}(c) (see Fig. S2 in Supporting Information for details on extraction of $R$-$t$ data points from images). The ICCD inset images indicate that, between 20 and 30~$\mu$s, the plume front is reflected by the substrate shield and propagates back toward the target. This supports the interpretation that the ion peak occurring 6.5~$\mu$s later arises from the reflected leading edge (I). This double-peak structure is well described by the superposition of incoming and reflected waveforms, each characterized by a sharp rise due to the leading edge arrival followed by an exponential-decay tail (Supporting Information, Fig.~S3). While the surface responsible for plume reflection here is the substrate shield---by the necessity of protecting the substrate during probe measurements---the same reflection effect is expected when the shield is replaced by the substrate heater during deposition. Both objects---shield and substrate heater---represent perpendicular, flat surfaces with similar plasma-exposed areas encountered by the expanding plume. The probe data between the first-pass and reflected signals imply a maximum instantaneous speed of 0.11~cm~$\mu$s$^{-1}$ for this structure at the shield/substrate location. 

ICCD images acquired at 50–200 $\mu$s delay times (Supporting Information, Fig. S1 and Movie S6) show the back-propagating plume front colliding with the target and subsequently propagating again toward the substrate. Measurements of the plume front as it traverses the field of view in both directions are consistent with the return of its leading edge to the ion probe at 186 $\mu$s with an instantaneous speed of 0.02~cm~$\mu$s$^{-1}$ (Supporting Information, Fig. S4). Hence, we attribute the late arrival probe peak (186 $\mu$s) to the plume front that has bounced between shield and target. It is essentially an attenuated replica of the peaks detected earlier at 22 and 28.5 $\mu$s. 

The sharp intensity peak (II) of the early expansion undergoes significant changes over time. Between 0.5 and 20 $\mu$s, its intensity maximum decreases by a factor of 20, and its FWHM broadens from 0.10 to 3.5 cm. Similar to the leading-edge feature, the peak position follows a quadratic drag progression, though in this case with a higher $\alpha = 1.17\ \mathrm{cm^{-1}}$, as shown in Fig. \ref{fig:ICCD}(c). This larger drag coefficient suggests either a greater collision cross section or propagation through a denser plume interior. Extrapolating this quadratic trend, the intensity maximum of feature II would reach the probe at 98 $\mu$s. If the feature consisted primarily of neutral species, however, it would not be detectable by the ion probe. As shown in Fig. \ref{fig:ICCD}(c) (inset), feature II encounters the reflected plume front while still en route to the substrate region. During this interaction, the position of the II intensity maximum appears to stall, and the image brightness increases as the front traverses the field of view. The transport and characteristics of the material in feature II are likely impacted by its interaction with the reflected and rebounded plume (see illustration in Fig. S4 of Supporting Information).

The third intensity maximum of the early plume (marked ``III'' in the Fig. \ref{fig:ICCD}(a) inset) cannot be tracked in the ICCD images past 2.4~$\mu$s due to low intensity. A quadratic drag fit to feature III data points (0--2.4 $\mu$s) yields $\mathrm{\alpha = 2.09\ cm^{-1}}$, reflecting its greater deceleration with respect to feature II (Supporting Information, Fig. S5). If it proceeded according to this quadratic drag trend, it would arrive at the probe with delays greater than 10 ms. Constituents of this component are also likely to interact with the reflected leading edge.

Overall, the ICCD and ion probe measurements underscore a semi-confined plasma synthesis medium with complex plume dynamics, that is amenable to control by adjustments in background pressure, laser irradiation parameters, pulse repetition rate, and geometry. Molecular cluster and nanoparticle formation are known to accompany plume expansions in this type of environment~\citep{Wood1998b,Geohegan1999, Geohegan1998,Mahjouri-Samani2017,Lin2020}. Hence, this setting is promising for engineering of porous metal chalcogenides assembled from plume-originated mesoscale building blocks. In this regard, it is interesting to first consider the potential effects of the reflection and rebounding of the plume front. Plume reflection due to interaction with a substrate heater or shield is well documented in PLD at relatively high background pressures (100--500~mTorr)~\cite{Geohegan1999}. The residue of reflected plumes has been observed to stagnate in the region between target and substrate allowing nanoparticle growth and aggregation for up to several seconds~\cite{Geohegan1999,Geohegan1998}. However, these slow-moving gas-phase objects are generally prevented from being incorporated into the growing films by thermophoretic forces arising from the gas temperature gradient established by the substrate heater. Furthermore, Rayleigh scattering studies in similar configurations suggest that, at our deposition temperature of $350\degree$C, pressures near 500~mTorr would be required to form appreciable number densities of proper nanoparticles (2–10~nm) in the stagnation zone~\cite{Geohegan1999,Geohegan1998}. Hence, it is unlikely that nanoparticles from a stagnant reflected plume are significant contributors to film growth in our setup. Accordingly, we assess that the most likely candidates for mesoscale building blocks in our 100 mTorr Ar environment are small molecular clusters formed from the gas phase during the initial plume traversal. These clusters, traveling at speeds $\le$~0.11~cm~$\mu$s$^{-1}$ ($\le$~0.50~eV/atom), can be further driven onto the growing film by the tail of the initial expansion or, possibly, by the rebounded plume, which still travels at 0.02~cm~$\mu$s$^{-1}$ (200~m~s$^{-1}$) when arriving at the substrate---a sufficiently high speed to promote the incorporation of clusters into the film. Evidently, it is possible that in addition to molecular clusters, a class of very small nanoparticles ($<$1--2 nm), formed in the quasi-stagnant intervening gas, could also be driven onto the substrate by the rebounded 0.02~cm~$\mu$s$^{-1}$ plume. 

Numerous other effects may be important in this semi-confined environment, since the species contributing to film growth may have a complex history of gas-phase interactions. For example, collisions between the charged leading edge---rebounding between the substrate and the target---and slower-moving plume components in the central and rear regions of the plume may play an important role in determining the size and ionization state of the film building blocks, as well as the efficiency of their incorporation. This interaction can probably be tuned during film growth by adjusting the pulse repetition rate, target-to-substrate distance, and background pressure. These may allow additional optimization of the porous mesostructure. 

Microstructural analysis of thin films produced by the decelerated and confined material flux can provide further insight into the nature of the film building blocks and plume constituents. We begin this analysis using XRR, which yields information on the density, porosity, roughness, and thickness of the films. 

\begin{figure}[!t]
    \centering    \includegraphics[width=0.7\linewidth]{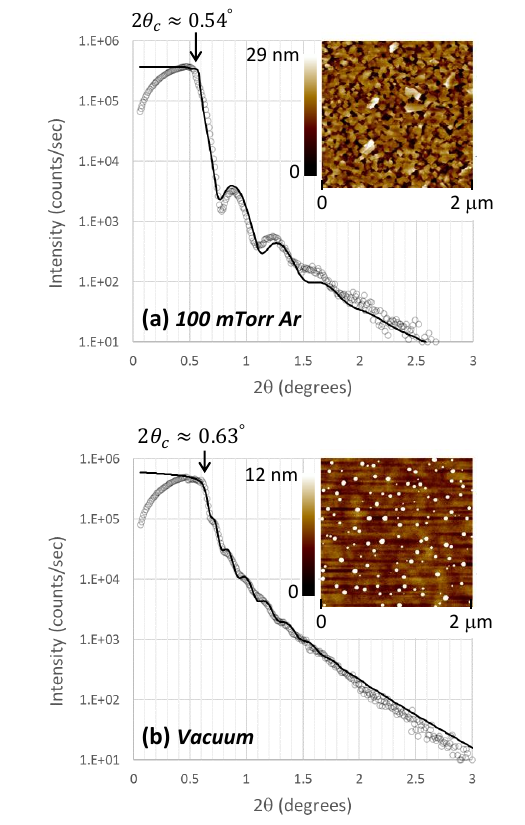}
    \caption{X-ray reflectivity scans (open circles) of films grown in (a) 100 mTorr Ar and (b) vacuum (1.2$\times10^{-6}$ Torr). Both films grown with 6,000 laser pulses under the same irradiation conditions: fluence = 1.0 J cm$^{-2}$; spot area = 7.5 mm$^{2}$. Three-layer model (air, film, substrate) fits (solid line) obtained with the differential evolution algorithm of GenX 3 using Parratt's reflectivity formula approximately describe the data. The film parameters corresponding to the fits are shown in Table \ref{tab:properties}. The insets exhibit atomic force microscopy scans, showing that the Ar background film comprises a mesoporous framework with voids of sizes below $\sim$100 nm. The vacuum film is smoother. Small circular features on the surface of the vacuum film $\mathrm{(\sim\!3.5\times 10^9 \ cm^{-2})}$ are most likely due to post-growth ambient exposure.
    }
    \label{fig:XRR}
\end{figure}

In XRR, the critical angle for X-ray transmission across the air/film interface is $\theta_c = \sqrt{2 \delta}$, where $\delta$ represents the deviation of the X-ray refractive index from unity and is proportional to the electron density of the film material~\citep{Birkholz2006}. Hence, $\theta_c$ scales with the mass density of the film. The angular separation between interference fringes is related to the film thickness, while the rate of damping of the reflected intensity with increasing angle is influenced by the roughness of the air/film and the film/substrate interfaces. The angular damping rate increases with rougher interfaces. These dependencies allow measurement of several film parameters.

X-ray reflectivity scans of films grown in 100 mTorr Ar background and in vacuum are shown in Fig. \ref{fig:XRR}. Both films were grown with 6,000 laser pulses under the same laser irradiation conditions (1.0 J cm$^{-2}$ fluence; 7.5 mm$^{2}$ spot area). The solid traces indicate fits to the data using Parratt's recursive reflectivity formula~\citep{Parratt1954} employing the differential evolution algorithm of GenX 3 (Ref.~\citenum{Glavic2022}). A three-layer structure was used, consisting of air (above the sample), a thin film, and a substrate. The XRR intensity of the Ar background sample is best described by modeling the film layer as a mixture of 85\% FeSe and 15\% voids, hence indicating a porous structure. This corresponds to an overall density of 4.76 g cm$^{-3}$, which is significantly lower than the density of bulk FeSe (5.70--5.72 g cm$^{-3}$,  Refs.~\citenum{mcqueen2009extreme,hota2024investigation}). The scattering length density (SLD) needed to compute the XRR intensity was obtained from SLD values of the constituents (bulk FeSe and air) using the standard volume fraction approach. The experimental XRR pattern of Fig. \ref{fig:XRR}(a) cannot be fitted by models where the film consists of only FeSe with the bulk density. Noteworthy is the steep drop in reflectivity between  $2\theta_c \approx 0.54^\circ$ and $2\theta_c \approx 0.75^\circ$, which can only be reproduced by models including pores. The lower density implied by the porous model agrees with the observation that the inert background film appears to show a lower critical angle (\( 2\theta_c \approx 0.54^\circ \)) than the vacuum film (\( 2\theta_c \approx 0.63^\circ \)). The porosity inferred from XRR is supported by the surface morphology revealed by AFM scans (Fig. \ref{fig:XRR} insets). The Ar background film has a root-mean-square (RMS) roughness of 5.3 nm with height variations of tens of nanometers, and a microstructure with void sizes below roughly 100 nm. The numerous sharp 90$^\circ$ corners noted by AFM are suggestive of the tetragonal motif of $\beta$-FeSe. The vacuum film is significantly smoother with RMS roughness of 0.56 nm, measured ignoring the small round features on the surface. Such features are most likely due to post-growth ambient exposure~\citep{Hiramatsu2019,Obata2021}. Judging from the surface morphology, the vacuum film appears compact with no sign of voids or porosity.

\begin{table}[!b]
\centering
\caption{Thin film structural and morphological parameters extracted by fitting Parratt's recursive formula~\cite{Parratt1954} for a three-layer model (air, thin film, substrate) to the X-ray reflectivity data of Fig.~\ref{fig:XRR}, using the differential evolution refinement of the GenX~3 software~\cite{Glavic2022}.}
\centering

\resizebox{0.6\linewidth}{!}{\begin{tabular}{@{}lllll@{}}
\toprule
\toprule
Physical Parameter & 100 mTorr(Ar)     & Vacuum \\ \midrule
Density (g cm$^{-3}$)  & 4.76   & 5.60   \\
FeSe fraction (\%)     & 85    & 100    \\
Void fraction (\%)     & 15    & --    \\
Thickness (nm)         & 21   & 40   \\
Top layer roughness (nm)  & 1.9    & 0.6   \\
Interface roughness (nm)  & 0.4    & 2.0   \\
\bottomrule \bottomrule
\end{tabular}}
\label{tab:properties}
\end{table}

Film thickness and interface roughness were also extracted from the XRR fits and are shown in Table \ref{tab:properties} along with density and porosity.  
We note that the growth rate in 100 mTorr Ar (0.0035 nm/pulse) is $\sim$2 times lower than in vacuum (0.0067 nm/pulse), indicating that the background gas prevents a substantial fraction of the ablated material from reaching the substrate. This resulted in an inert background film that is significantly thinner (21 nm) than the film grown in vacuum (40 nm). Also noteworthy are the roughness parameters inferred from XRR. Our reflectivity model yields a top layer roughness of 1.9 nm for the porous film, compared to a very smooth surface for the film grown in vacuum (0.6 nm). This is consistent with the trend of RMS roughness seen in the AFM scans, which yielded 5.3 nm and 0.56 nm for the porous and vacuum films, respectively. 

Also of interest are the roughness values for the FeSe/MgO interface. This interface is smooth (0.4 nm) for the Ar background film. This low roughness reflects the pristine, as-prepared MgO surface. As revealed by the combined analysis of ion probe and ICCD data, the porous film grows under low energy ion irradiation ($\le$~0.50~eV/atom), which preserves the original MgO surface. The higher interface roughness of the vacuum film (2.0 nm) indicates that processes such as sputtering, ion implantation, and defect formation due to displacement damage were significant. Indeed, film growth in vacuum occurred under ion irradiation with distribution centered at $\sim$100~eV/atom. 

\begin{figure*}[!t]
    \centering
    \includegraphics[width=1.0\linewidth]{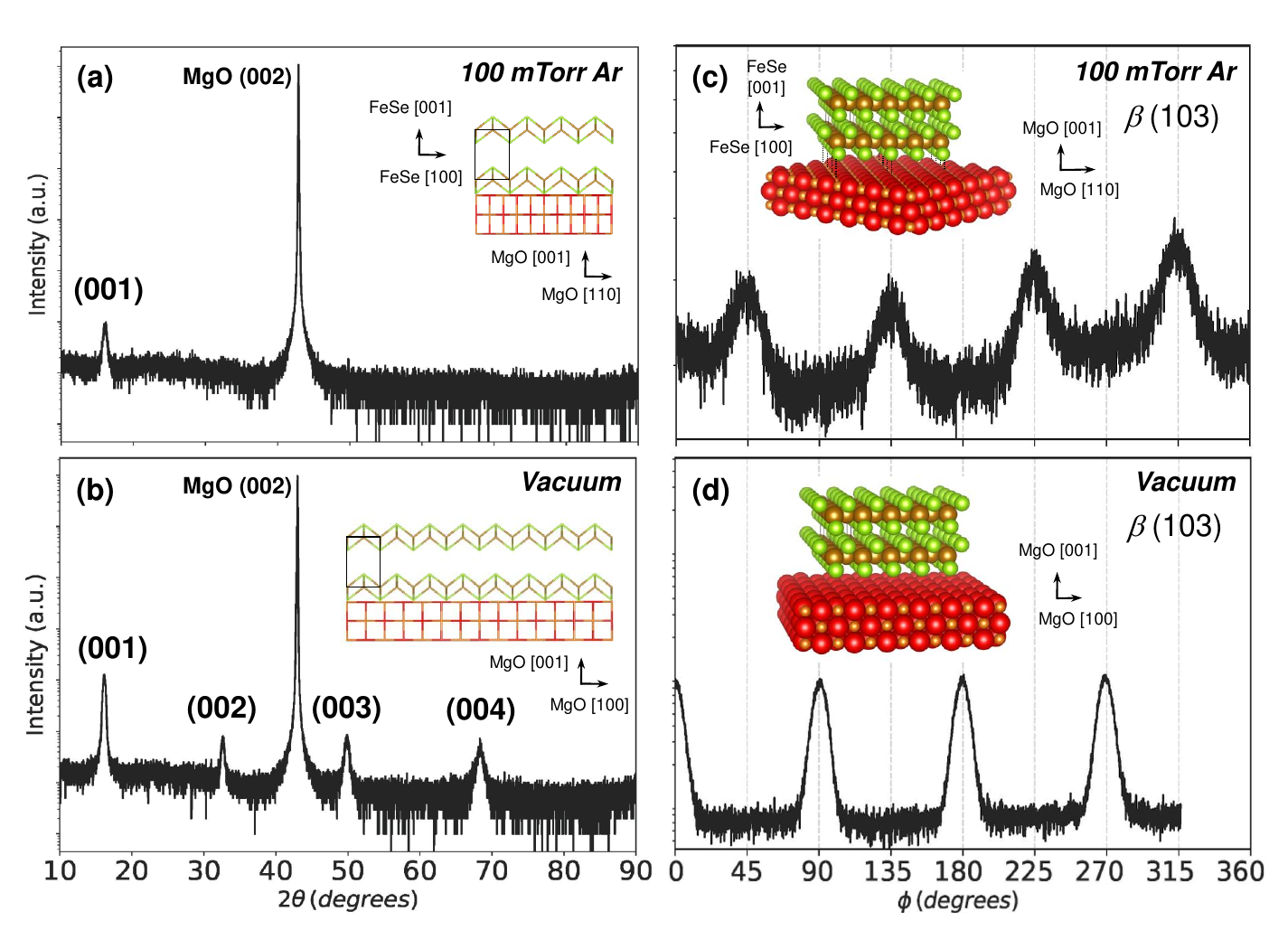}
    \caption{XRD $\theta$-2$\theta$ scans of films grown in (a) 100 mTorr Ar and (b) vacuum, showing reflections corresponding to $c$-axis oriented $\beta$-FeSe. High order reflections are absent in the Ar background film, implying lower long-range crystal coherence. XRD $\phi$ scans of the $\beta$-FeSe (1~0~3) reflection for both films (c,d) display 4-fold symmetry, indicating in-plane epitaxy with respect to MgO. The (c) 100 mTorr film is oriented such that \mbox{$\beta$-FeSe~[1~0~0]~ $\parallel$~[1~1~0]~MgO} (as illustrated in the inset) while the vacuum film grows with \mbox{$\beta$-FeSe [1 0 0] $\parallel$ [1 0 0] MgO}. The insets in panels (a) and (b) show possible domain-matching epitaxial relationships for the two orientations. Fe and Se atoms are shown as brown and green spheres or vertices; O and Mg in the MgO substrate are depicted in red and brown, respectively.}
    \label{fig:XRD}
\end{figure*}

Figure \ref{fig:XRD}(a) shows a symmetric $\theta$-2$\theta$ XRD scan for the film grown in 100 mTorr Ar. The crystal structure of the mesoporous film is clearly the tetragonal $\beta$-FeSe phase, as indicated by the (0 0 1) reflection at 2$\theta = 16.17^\circ$. This implies that the mesoporous film is $c$-axis oriented with the epitaxial relation $\beta$-FeSe (0 0 1)$\parallel$(0 0 1) MgO. The lattice parameter of the tetragonal lattice in the direction normal to the surface is found to be $c$ = 5.485~\AA, based on a calibration of the angular scale using the high-intensity MgO (0 0 2) substrate peak located at $2\theta = 42.96^\circ$. The mesoporous film is assumed to be relaxed, given its open structure, which is unlikely to be able to hold stress. The $c$ parameter is known to depend on the exact stoichiometry of $\beta$-FeSe. The value inferred from our scan is consistent with Fe-rich $\beta$-FeSe thin films previously reported~\cite{Hsu2008,Obata2021}. For comparison, Fig. \ref{fig:XRD}(b) displays the \mbox{$\theta$-2$\theta$} scan for a film grown in vacuum. Out-of-plane epitaxy, with the $c$-axis of the tetragonal $\beta$-FeSe structure aligned with the direction normal to the substrate is also observed, with the same $c$ = 5.485 \AA{}  parameter, likely indicating a relaxed film as well. In the vacuum film, the (0 0 1) peak is accompanied by the (0 0 2), (0 0 3), and (0 0 4) reflections. These higher order reflections are absent in the porous layer. Hence, the Ar background film shows significantly lower long-range crystalline coherence compared to the film grown in vacuum. 

The in-plane crystallographic orientation of the films can shed light into the origin of their distinct long-range order characteristics. To probe the in-plane orientation, we performed $\phi$ scans of the (1 0 3) asymmetric reflection of $\beta$-FeSe. The $\phi$ scan for the mesoporous film is shown in Fig. \ref{fig:XRD}(c). The four-fold symmetry reveals that the mesoporous film is also epitaxial with respect to the substrate in the in-plane directions. We note that the base of the $\beta$-FeSe unit cell is rotated by 45$\degree$ with respect to the (1 0 0) direction of MgO, which was set to 0$\degree$ when setting up the $\phi$ scan. This indicates that the in-plane orientation of the Ar background film is such that \mbox{$\beta$-FeSe~[1~0~0]~ $\parallel$~[1~1~0]~MgO}, as illustrated in the inset of the figure. This contrasts with the growth habit of the vacuum film.  In the latter, the $\phi$ scan in Fig. \ref{fig:XRD}(d) reveals in-plane orientation with \mbox{$\beta$-FeSe [1 0 0] $\parallel$ [1 0 0] MgO}. The sides of the square base of the FeSe unit cell are parallel to the sides of the square face of the MgO unit cell---``square-on-square'' epitaxy for brevity.

These two different in-plane orientations have been observed in numerous experiments of PLD-grown $\beta$-FeSe on MgO~\cite{Wu2009,Obata2021}. They are often present simultaneously in the same sample and their relative fraction is strongly dependent on the growth temperature. The fact that the FeSe porous film is fully 45-degree oriented with respect to the MgO lattice---while the vacuum film shows only square-on-square orientation, even though both were grown at the same temperature---suggests that, in the present case, the different orientations arise from distinct crystal growth mechanisms. As inferred from the ICCD and ion probe measurements, the growth of the mesoporous film is dominated by attachment of pre-formed clusters or very small nanoparticles, further driven onto the film by the tail of the initial expansion or by the rebounded plume~\citep{Geohegan1999,Geohegan1998,Mahjouri-Samani2017,Lin2020}. This is consistent with the porous framework revealed by XRR and AFM (Fig. \ref{fig:XRR}). The low kinetic energy of these mesoscale building blocks ($\le$0.50~eV/atom) preserves the order of the MgO surface. As illustrated in the inset of Fig. \ref{fig:XRD}(c), the MgO surface---which essentially does not reconstruct~\cite{Urano1983,Barth2003}---has corrugations formed by rows of Mg atoms along the (1~1~0) directions of the crystal structure. Judging from STM measurements and DFT calculations of vacuum-exposed $\beta$-FeSe surfaces~\cite{Liu2012,Song2012,Huang2016}, small crystallites formed in the gas phase are likely selenium(Se)-terminated. Hence, the surface of $\beta$-FeSe crystallites probably features nascent rows of (1~0~0)-oriented Se atoms. These protruding Se rows~\cite{Huang2016} approximately fit along the diagonal depressions occupied by the Mg atoms in MgO. A low-free-energy configuration is expected when the Se rows settle into the MgO corrugations upon the soft landing of crystallites. This arrangement also creates the favored low energy chemical coordination between anions and cations across the $\beta$-FeSe/MgO interface (i.e., Mg coordinates with Se; Fe coordinates with O). This interface ordering is, of course, far from perfect due to the mismatch of $\sim$20\% between the inter-row distances of Se protrusions and the Mg depressions. This results in a first layer of crystallites that are tilted by small angles with respect to the substrate. Hence, although epitaxial, the mesoporous film is predicted to have a lower degree of atomic periodicity and long-range order. Mechanisms for atomic plane reorganization, such as epitaxial relaxation, dislocation glide, and surface diffusion, are bound to be less efficient as the film grows by cluster or nanoparticle accretion, compared to conventional epitaxial growth from atomic or molecular species. It should also be mentioned that high roughness of the top layer of the mesoporous film---evident in both AFM and XRR data---imply further weakening of any higher-order Bragg reflections. Hence, the mosaicity induced by crystallite misorientation upon attachment and the rough mesoscale surface are consistent with the presence of only the first-order $\theta$-2$\theta$ reflection in the Ar background film. Any low intensity higher-order peaks are undetectable given the signal-to-noise ratio of the low thickness and density of the mesoporous film. 

In vacuum, UV nanosecond PLD is known to produce plumes rich in atomic species. This is attested by a large body of PLD experiments that show layer-by-layer film growth from adatom attachment, including for $\beta$-FeSe on MgO~\cite{Obata2021}. In our case, this behavior is evidenced by the intense peak observed in the ion probe data in vacuum, attributed to singly charged atomic species en route to the substrate with kinetic energy $>$100~eV/atom. It has also been established that an Fe-rich layer (1--2 nm) is present at the FeSe/MgO interface. This layer is found below the position of the original MgO surface, indicating it forms by diffusion of Fe atoms into the substrate~\cite{Obata2021}, which in our case is certainly enhanced due to high-energy ion irradiation~\cite{Metev1994}. The Fe-rich slab acts as an accommodation layer that supports the square-on-square orientation of FeSe on MgO. In fact, the deliberate introduction of an Fe buffer layer in the process (i.e., FeSe/Fe/MgO) enables control of epitaxial phenomena and can be used to switch the FeSe growth between domain-matching and lattice-matching epitaxy~\cite{Obata2021}. When homogenized by relatively high temperature deposition ($T\geq350\degree$C), the registry of this accommodation layer favors the FeSe [1~0~0]~ $\parallel$~[1~0~0] MgO orientation~\cite{feng2018tunable,harris2019double,Obata2021}. Strain due to lattice mismatch is relaxed via three different domain-matching epitaxy relationships (11/10; 10/9; 9/8)~\cite{ Obata2021}. This is illustrated in the inset of Fig. \ref{fig:XRD}(b). As noted in Table \ref{tab:properties}, the $\beta$-FeSe/MgO interface of our vacuum film yields an XRR roughness parameter of 2.0 nm, again attributed to energetic ion irradiation. This level of roughness suggests that the interface layer could extend across approximately four FeSe unit cells. Although atomically sharp FeSe/MgO interfaces have been observed on PLD films, in the case of our vacuum film the lattice diagram in Fig. \ref{fig:XRD}(b) serves only as an illustration of a possible domain-matching relationship between the structures.

\section{Conclusions}
\label{sec4}
We have applied PLD to the synthesis of mesoporous thin films of the metal chalcogenide $\beta$-FeSe on MgO. Growth of the porous layer was carried out in inert Ar background using a UV-nanosecond-laser-generated plasma plume produced by the ablation of an FeSe target. Imaging and ion probe measurements in 100 mTorr Ar showed that the initial plume consists of three main structures with distinct rates of expansion. These plume components interact with the substrate heater and the ablation target, giving rise to complex spatiotemporal dynamics involving collisions between the rebounding charged leading edge and slower-moving species in the central and rear regions of the plume. This environment represents a semi-confined novel reactor geometry that affords great variability for engineering the porosity and crystal structure of cluster and nanoparticle-based metal chalcogenide electrocatalyst films. XRR and AFM on films grown at 350$\degree$C revealed 15\% porosity and pore sizes below $\sim$100 nm. The porous framework likely formed by oriented attachment of gas-phase clusters or very small nanoparticles with kinetic energy $\le$0.50 eV/atom, further driven onto the film by the tail of the initial expansion or the rebounded plume. The porous films are epitaxial with the $c$-axis of $\beta$-FeSe oriented normal to the substrate, while the in-plane orientation invariably has the square base of the $\beta$-FeSe unit cell rotated by 45$\degree$ with respect to the MgO cell. We suggest this results from soft landing of pre-formed FeSe crystallites on the MgO substrates, where protruding Se rows of $\beta$-FeSe settle onto (1~1~0)-oriented corrugations of the MgO surface. Our study implies that synthesis of candidate electrocatalyst materials by PLD in inert background may enable mesoporous frameworks that have a single crystallographic orientation and uniform/homogeneous crystal facets available for electrochemical reactions and active surface investigations. An integrated approach---combining ICCD and ion probe diagnostics, modeling, and feedback from correlated property characterization---can drive the refinement of this process to yield optimal materials performance. Further studies on porosity/pore size control, the impact of substrate choice on in-plane orientation, the effect of experimental parameters on the dynamics of the rebounding plume, and the nature of the gas-phase building blocks (e.g., using matrix-assisted laser desorption/ionization and molecular dynamics simulations) are needed to fully assess the applicability of this growth method in electrocatalyst engineering and confirm growth mechanisms.

\section*{Acknowledgments}
This work was supported in part by the National Science Foundation (NSF) EPSCoR RII-Track-1 Cooperative Agreement OIA-2148653. Plasma measurements as well as thin film growth and characterization was supported by the Center for Nanophase Materials Sciences (CNMS), which is a US Department of Energy, Office of Science User Facility at Oak Ridge National Laboratory.

\bibliographystyle{elsarticle-num} 
\bibliography{bibliography.bib}






\end{document}